\documentclass[review]{elsarticle}

\usepackage{lineno,hyperref}
\usepackage{amsfonts}
\usepackage{amsmath,amssymb}
\usepackage{multirow}
\usepackage{xcolor}
\modulolinenumbers[5]

\journal{ }

\begin{document}

\begin{frontmatter}
\linenumbers

\title{Non-relativistic solutions for three-body molecules within a Chern-Simons model}


\author[mymainaddress]{Francisco Caruso}\corref{mycorrespondingauthor}
\cortext[mycorrespondingauthor]{Corresponding author}
\ead{francisco.caruso@gmail.com or caruso@cbpf.br}

\author[mysecondaryaddress]{Vitor Oguri}
\author[mysecondaryaddress]{Felipe Silveira}
\author[mymainaddress]{Amos Troper}

\address[mymainaddress]{Centro Brasileiro de Pesquisas F\'{\i}sicas -- Rua Dr.~Xavier Sigaud, 150, 22290-180, Urca, Rio de Janeiro, RJ, Brazil}
\address[mysecondaryaddress]{Instituto de F\'{\i}sica Armando Dias Tavares, Universidade do Estado do Rio de Janeiro -- Rua S\~ao Francisco Xavier, 524, 20550-900, Maracan\~a, Rio de Janeiro, RJ, Brazil}

\begin{abstract}
The wave functions and the ground state energies for the bound states of four different muonic and electronic molecules, governed by the Chern-Simons potential in two spatial dimensions, are numerically obtained with the Numerov method. The new results are compared with former planar configuration studies that consider the attractive potential as being proportional to $\ln(\rho)$, as well as with the three-dimensional analogs assuming a Coulomb potential.
\end{abstract}


\end{frontmatter}

\newpage

\section{Introduction} \label{intro}

The interest in the Chern-Simons theory transcended its original domain of application~\cite{Witten, Hu, CS-Deser1, CS-Deser2}. This trend is due in part to the understanding that there is a formal connection between planar quantum electrodynamics and the Chern-Simons theory~\cite{Marino}. Indeed, it has been applied to describe different effects and systems: the Quantum Hall Effects~\cite{Klitzing, Tsui, Zhang, Read, Zhang-book}, when it was argued that the scaling limit of a quantum Hall system is described by quantum Chern-Simons theory, thereby explaining the discreteness of the set of values of the Hall conductivity~\cite{Frohlich}; the electron-electron interaction in planar QED~\cite{Yang, Guo, Girotti, Magalhaes}; the hydrogen atom~\cite{Caruso}; and, finally, in the description of muonic and electronic atoms, with nuclei composed of a proton, deuteron, and Triton in Atomic Physics~\cite{felipe5}.

These last results are based on the so called Maxwell-Chern-Simons theory which is a topological mass theory~\cite{Paschoal, Boyanovsky, Dunne}. The present work is designed to develop a theoretical model for a molecule composed of three particles (similar to the ionized hydrogen molecule) assuming their electromagnetic interactions to be governed by a Chern-Simons potential, following what was recently done for atoms involving the same particles~\cite{felipe5}.

To carry on this line, we need first to know the ground state wave function of the atoms that make up the studied molecules, which have already been obtained in a recent work~\cite{felipe5}. This result is summarized in Section~2, where analytical expressions for each wave function are obtained by fitting the numerical solutions. In Section~3, the two dimensional time-independent Schr\"{o}dinger equation, which describes a three-body quantum system formed by two positive nuclei (in this work we will use just protons and deuterons) and one negative particle (electron or muon) is determined in an adiabatic approximation. Doing so, the usual $W_\pm$ contributions to the effective molecular potential are calculated in Section~4, which allow us to obtain the interatomic potential of each molecule, following the steps of the references~\cite{park, felipeepl, felipeepjd} and mainly that of Ref.~\cite{felipe5}. And finally, in Section~5, the resulting Schr\"{o}dinger equation will be solved numerically by applying the Numerov method. The calculated ground state energies and respective eigenfunctions are given in this Section. Also, the mean distance between the two nuclei inside each molecule is calculated. A comparison is made with results from different models. Some final remarks are given in Section~6.

\section{2D Atom in a nutshell}

In order to build our molecular adiabatic model, we need to know an analytical expression for the wave functions of the atoms that make up such a molecule. To determine such an analytical expression for the ground state wave function of the two-dimensional atom, we will use the results obtained in the reference~\cite{felipe5}.

The numerical solutions of equation (6) of the reference~\cite{felipe5} can be used to plot the desired wave functions. Then, using the Mathematica 12.1 software, we are able to fit these functions, getting the analytical expressions we need.

\begin{figure}[htbp]\label{fit}
\centerline{\includegraphics[width=8cm]{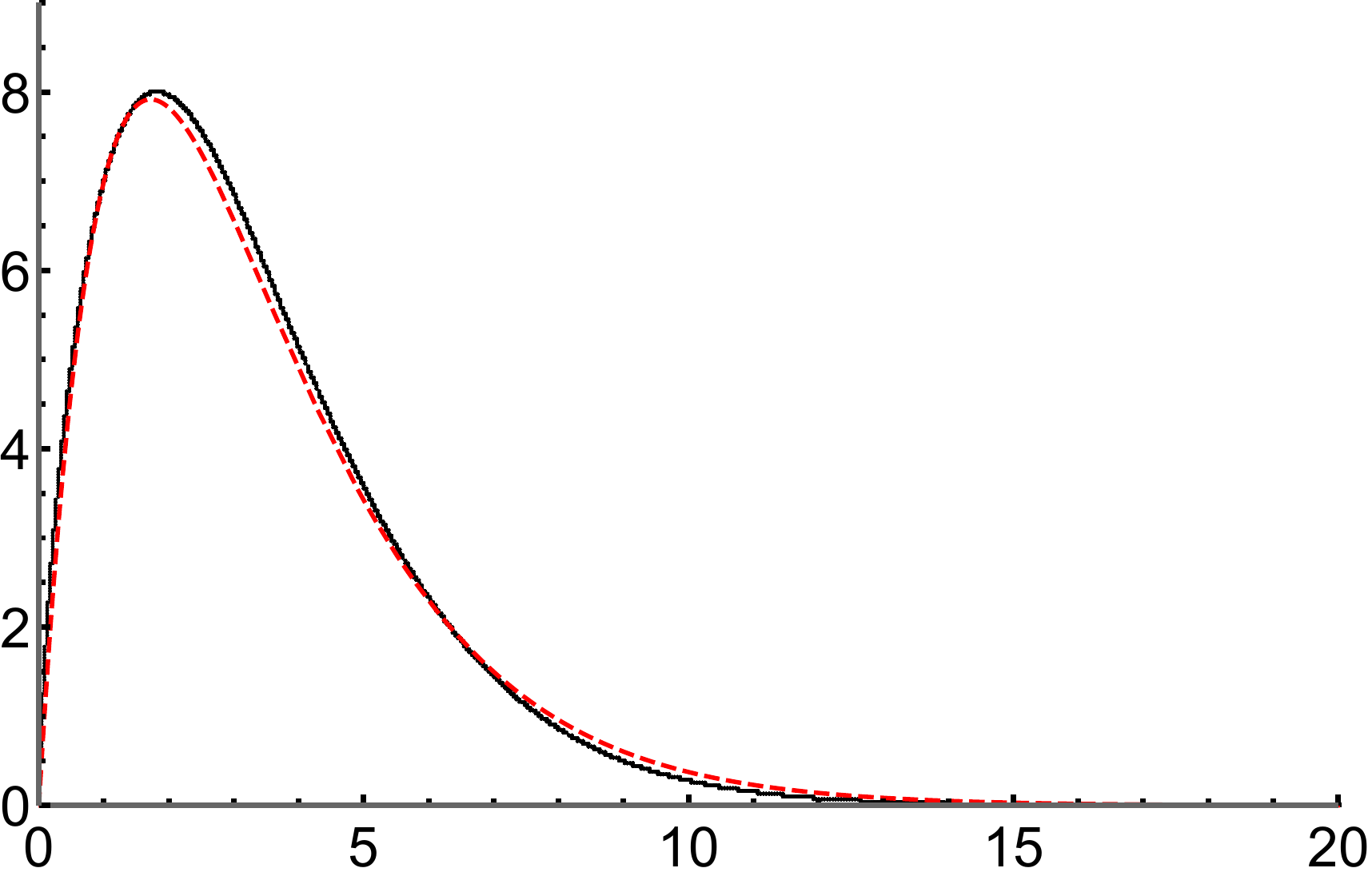}}
\caption{Fit of the ground state wave function of the hydrogen atom (dotted line) with energy $-2.2417$ and wave function $u(r) = 12.8453 r e^{-r/1.72068}$.}
\end{figure}

In figure~\ref{fit}, we can see an example of a fit made to obtain the $u(r)$ wave function for the $pe$ atom. For simplicity, we are not going to reproduce the fit of all wave functions.

So we can easily get the $\psi$ wave function from this result. With $\psi(r) =A u(r) r^{-1/2}$, being $A$ the normalization constant, we have

$$\psi(r) = 12.8453\, A\, r^{1/2} e^{-r/1.72068} $$

\noindent The normalization is fixed from the equality

$$A^2 \int_{0}^{\infty} \left(\frac{u(r)}{\sqrt{r}}\right)^2 2\pi r\,\mbox{d}r = 2\pi A^2 (12.8453)^2 \int_{0}^{\infty} (e^{-r/1.72068})^2 r^2\, \mbox{d}r = 1 $$

\noindent Therefore,

$$A = \frac{1}{\sqrt{1320.04}} = 0.0275237 $$
and, then, for the $pe$ atom, the wave function $\psi(r)$ is given by

$$\psi(r) = 0.353501 \sqrt{r}\, e^{-r/1.72068}$$

\noindent Making $r_i = \rho_i \rho_0$,\footnote{comparing with Ref.~\cite{felipeepjd}, $\rho_{a} = \rho_{13}$ e $\rho_{b} = \rho_{23}$.}

$$\psi_{13} =  0.353501 \sqrt{\rho_{13}\rho_0}\, e^{-\rho_{13}\rho_0/1.72068}$$
$$\psi_{23} = 0.353501 \sqrt{\rho_{23}\rho_0}\, e^{-\rho_{23}\rho_0/1.72068}$$

The generic (non-normalized) ground state wave function for all atoms used in this work is of the form $\psi(r) = a \sqrt{r}\, e^{-r/b}$, where the parameters, $a$ and $b$, can be adjusted for each atom and the results are displayed in table~\ref{atomos-par}.
\renewcommand{\arraystretch}{1.2}
\begin{table}[ht]
\caption{Parameters $a$ and $b$ of the ground state wave functions for different atoms and different values of the Chern-Simons parameter $\lambda$ (See Section~3)}
\begin{center}
\begin{tabular}{|c|c|c|c|c|}
    \hline
    &\multicolumn{2}{c|}{$\lambda=0.2\times10^{-3}$}&\multicolumn{2}{c|}{$\lambda=0.2\times10^{-5}$}\\ \hline
    Atom         &  $a$      &  $b$     &     $a$     &     $b$          \\ \hline
      $pe$       & 12.6765   & 1.74095  &   12.8453   &   1.72068        \\ \hline
      $de$       & 12.6772   & 1.74082  &   12.7106   &   1.7383         \\ \hline
      $p\mu$     & 12.8302   & 1.72352  &   12.8298   &   1.71609        \\ \hline
      $d\mu$     & 12.8373   & 1.71479  &   12.8285   &   1.71633        \\ \hline
\end{tabular}
\end{center}
\label{atomos-par}
\end{table}
\renewcommand{\arraystretch}{1}

\newpage
\section{2D Molecule}

Following the steps of references~\cite{park, felipeepl, felipeepjd}, the planar time-independent Schr\"{o}\-dinger equation describing a three-body quantum system formed by two positive nuclei (in this work we will use just protons and deuterons) and one negative particle (electron or muon) orbiting them is:
\begin{eqnarray}\label{eq-molecula}
  &\,&\left[-\frac{\hbar^2}{2m_1}\nabla^2_{1}-\frac{\hbar^2}{2m_2}\nabla^2_{2}-\frac{\hbar^2}{2m_3}\nabla^2_{3}
+V(\vec{r}_{13},\vec{r}_{23},\vec{r}_{12})\right]\Psi(\vec{r}_{13},\vec{r}_{23},\vec{r}_{12}) = \, \nonumber \\
   && \quad = E_T\Psi(\vec{r}_{13},\vec{r}_{23},\vec{r}_{12})
\end{eqnarray}
with 1 and 2 being the indices of the two positive nuclei and 3 denotes the negative particle, $\vec{r}_{13} =\vec{r}_3-\vec{r}_1 $, $\vec{r}_{23}=\vec{r}_3-\vec{r}_2$ and $\vec{r}_{12}=\vec{r}_2-\vec{r}_1$.

The Chern-Simons potential of mutual interactions, with $r_{i} = |\vec{r}_{i}|$, is
\begin{eqnarray}\label{eq-molecula-electron}
  V(\vec{r}_{13},\vec{r}_{23},\vec{r}_{12})&=& -\frac{Z_1 e^2V_0}{2\pi}K_0\left(\frac{m_{\gamma}c}{\hbar}r_{13}\right) + \\
  &-& \frac{Z_2 e^2V_0}{2\pi}K_0\left(\frac{m_{\gamma}c}{\hbar}r_{23}\right) + \frac{Z_1 Z_2 e^2V_0}{2\pi}K_0\left(\frac{m_{\gamma}c}{\hbar}r_{12}\right) \nonumber
\end{eqnarray}
\noindent Here, $m_\gamma$ is the photon effective topological mass which will be expressed in terms of the electron mass through the relation $m_\gamma = \lambda m_e$. The value of $m_\gamma$ is not yet numerically predicted. It is expected to be in the range $0.1 < m_\gamma < 20$~eV. In order to be able to compare pour results with previous one we will investigate in this paper the range $0.2 \times 10^{-5} < \lambda < 0.2 \times 10^{-3}$ as done in Ref.~\cite{felipe5}. Throughout the paper predictions will be made for the two $\lambda$ threshold values.

We can rewrite the Eq.~(\ref{eq-molecula}) in relation to the center of mass of the particles that form the nucleus with $\Psi(\vec{R},\vec{r}_{12})=\chi(\vec{R})\psi(\vec{r}_{12})$, obtaining

\begin{equation}\label{22}
  -\frac{\hbar^2}{2M}\nabla^2_{_R}\, \chi(\vec{R})=E_{_{CM}}\,\chi(\vec{R})
\end{equation}
\begin{equation}\label{23}
  \left[-\frac{\hbar^2}{2\mu}\nabla^2_{r_{12}}-\frac{\hbar^2}{2m_3}\nabla^2_{\vec{r}_3}+V(r)\right]\psi(r)=E\psi(r)
\end{equation}

\noindent Since the whole dynamics of the problem is contained in the Eq.~(\ref{23}), we finally have to solve

\begin{eqnarray}\label{24}
  &&\left[-\frac{\hbar^2}{2\mu}\nabla^2_{r_{12}}-\frac{\hbar^2}{2m_3}\nabla^2_{\vec{r}_3}-\frac{Z_1 e^2V_0}{2\pi}K_0\left(\frac{m_{\gamma}c}{\hbar}r_{13}\right)\right.+\nonumber \\
  &&\left.- \frac{Z_2 e^2V_0}{2\pi}K_0\left(\frac{m_{\gamma}c}{\hbar}r_{23}\right) + \frac{Z_1 Z_2 e^2V_0}{2\pi}K_0\left(\frac{m_{\gamma}c}{\hbar}r_{12}\right)\right]\psi(r)=E\psi(r)
\end{eqnarray}

\noindent Making the transformations, $m_\gamma = \lambda m_e$, $\mu =\zeta m_e$, $r_{12} = \frac{\rho_{12} \rho_0}{\sqrt{\zeta}}$, $r_{13} = \rho_{13} \rho_0$, $r_{23} = \rho_{23} \rho_0$, $r_3 = \rho_{3} \rho_0$, $\rho_0 = \sqrt{\frac{\hbar^2}{m_e e^2 V_0}}$ and considering the Hartree atomic units,  $\hbar = m_e = e = 1$, $c = \frac{1}{\alpha}=137.0356$ and $V_0 = 1$, we have

\begin{eqnarray}\label{25}
  &&\left[-\frac{1}{2}\nabla^2_{\rho_{12}}-\frac{1}{2m_3}\nabla^2_{\rho_3}-\frac{Z_1}{2\pi}K_0\left(\frac{\lambda}{\alpha}\rho_{13}\right)\right.+\nonumber \\
  &&\left.- \frac{Z_2}{2\pi}K_0\left(\frac{\lambda}{\alpha}\rho_{23}\right) + \frac{Z_1 Z_2}{2\pi}K_0\left(\frac{\lambda}{\alpha}\frac{\rho_{12}}{\sqrt{\zeta}}\right)\right]\psi=E\psi
\end{eqnarray}

\noindent Regrouping the terms, we have
\begin{eqnarray}\label{27}
  \left\{-\nabla^2_{\rho_{12}} + \frac{1}{m_3}\left[\frac{Z_1 Z_2 m_3}{ \pi}K_0\left(\frac{\lambda}{\alpha}\frac{\rho_{12}}{\sqrt{\zeta}}\right)+W_\pm\right]\right\}\psi = \varepsilon \psi
\end{eqnarray}

\noindent with $\varepsilon = 2E$ and, in the adiabatic approximation, $W_\pm$ is the expected value for the energy $\varepsilon_e$ of the equation that describes only the electron movement, when we consider the fixed nuclei
\begin{equation}\label{ee}
\varepsilon_e \psi= \left[-\nabla^2_{\rho_3}-\frac{Z_1 m_3}{\pi} K_0\left(\frac{\lambda}{\alpha}\rho_{13}\right)-\frac{Z_2 m_3}{ \pi} K_0\left(\frac{\lambda}{\alpha}\rho_{23}\right)\right] \psi
\end{equation}

\noindent For simplicity, from now on we will call $\rho_{12} = \rho$, then, finally we can rewrite the Eq.~(\ref{27}) using $\psi(\rho,\theta)= \frac{u(\rho)}{\sqrt{\rho}}e^{\pm i \ell \theta}$ as
\begin{eqnarray}\label{28}
 \left\{\frac{\mbox{d}^2}{\mbox{d}\rho^2}+\varepsilon - \frac{1}{m_3}\left[\frac{Z_1 Z_2 m_3}{\pi}K_0\left(\frac{\lambda}{\alpha}\frac{\rho}{\sqrt{\zeta}}\right)+W_\pm \right]-\frac{\ell^2-0.25}{\rho^2}\right\}u(\rho) = 0
\end{eqnarray}

\section{The $W_\pm$ contribution to the effective molecular potential}

In order to numerically solve the Eq.~(\ref{28}), we first need to find the expression for $W_\pm$, following the steps of the references~\cite{felipeepjd, park}. The functions $W_\pm$ are given by the general expression

\begin{equation}\label{w+-}
  W_\pm = |N_\pm|^2 \int (\psi_{13} \pm \psi_{23}) \varepsilon_e (\psi_{13} \pm \psi_{23})\, \mbox{d}^{2D}V
\end{equation}

\noindent Substituting here the value of $\varepsilon_e$ given by Eq.~(\ref{ee}), we obtain

\begin{eqnarray}\label{w+-6}
  W_\pm &=& \eta + |N_\pm|^2 \int \left[ - \frac{m_3}{\pi} K_0\left(\frac{\lambda}{\alpha}\rho_{23}\right)\psi^2_{13}  -  \frac{m_3}{\pi} K_0\left(\frac{\lambda}{\alpha}\rho_{13}\right)\psi^2_{23}\right]\mbox{d}^{2D}V + \nonumber \\
  &\mp& |N_\pm|^2 \int \left[ \frac{m_3}{\pi} K_0\left(\frac{\lambda}{\alpha}\rho_{23}\right)\psi_{13} \psi_{23} + \frac{m_3}{ \pi} K_0\left(\frac{\lambda}{\alpha}\rho_{13}\right)\psi_{23} \psi_{13} \right] \mbox{d}^{2D}V \nonumber
\end{eqnarray}

\noindent Where $\eta$ is the ground state energy of one of the atoms that composes the molecule ($\eta$ values can be found in table~4 of reference~\cite{felipe5}). The expression for $W_\pm$ is usually written in terms of $\eta$ and three other functions ($\mathfrak{D}$, the direct integral, $\mathfrak{E}$, the exchange integral, and the overlap integral $\Delta$) defined, as usual, as:

$$W_\pm = \eta + \frac{\mathfrak{D}\pm\mathfrak{E}}{1\pm\Delta}$$

\noindent where

$$1\pm \Delta = \frac{1}{2|N_\pm|^2} = 2 \int |\psi_{13} + \psi_{23}|^2 \mbox{d}^{2D}V $$

$$\mathfrak{D} = \int \left[ - \frac{m_3}{\pi} K_0\left(\frac{\lambda}{\alpha}\rho_{23}\right)\psi^2_{13}  -  \frac{m_3}{\pi} K_0\left(\frac{\lambda}{\alpha}\rho_{13}\right)\psi^2_{23}\right]\mbox{d}^{2D}V $$

$$\mathfrak{E} = \int \left[ \frac{m_3}{\pi} K_0\left(\frac{\lambda}{\alpha}\rho_{23}\right)\psi_{13} \psi_{23} + \frac{m_3}{\pi} K_0\left(\frac{\lambda}{\alpha}\rho_{13}\right)\psi_{23} \psi_{13} \right] \mbox{d}^{2D}V$$

Thus, using the $\psi_{13}$ and $\psi_{23}$ values calculated in section 2, we are able to solve all the three equations above and therefore to compute $W_\pm$.

As a general comment, for $\ell=0$ states, we can say that all the potentials have very similar shapes for all the molecules considered here. They differ not so much on the height and width of the well. They depends also on the choice of the $\lambda$ factor.

\section{Numerical solutions}

Now that we know the expression of the $W_{\pm}$, we can rewrite Eq.~(\ref{28}) as

\begin{eqnarray}\label{282}
 \left\{\frac{\mbox{d}^2}{\mbox{d}\rho^2}+\varepsilon - U(\rho) \right\}u(\rho) = 0
\end{eqnarray}

\noindent with
\begin{equation}\label{potential28}
  U(\rho) =  \frac{1}{m_3}\left[\frac{Z_1 Z_2 m_3}{\pi}K_0\left(\frac{\lambda}{\alpha}\frac{\rho_{12}}{\sqrt{\zeta}}\right)+W_\pm \right] + \frac{\ell^2-0.25}{\rho^2}
\end{equation}

\renewcommand{\arraystretch}{1.2}
\begin{table}[htb]
  \caption{$\zeta$ and $m_3$ values for different molecules}
  \begin{center}
  \begin{tabular}{|c|c|c|}
    \hline
      Molecule     & $\zeta$           & $m_3$           \\ \hline
      $ppe$        &   918.076336715   &   1             \\ \hline
      $dde$        &   1835.24148394   &   1             \\ \hline
      $pp\mu$      &   918.076336715   &   206.7682830   \\ \hline
      $dd\mu$      &   1835.24148394   &   206.7682830   \\ \hline
  \end{tabular}
  \end{center}
  \label{tabelazeta}
\end{table}
\renewcommand{\arraystretch}{1}
For simplicity, only the solutions corresponding to $W_+$ will be discussed in this paper.\footnote{\, This choice can also be justified since, in 3D, only the $W_+$ term produces a potential well.} This potential can be adjusted for any molecule that is alike the ionized hydrogen molecule. Table~\ref{tabelazeta} gives the different values of the $\zeta$ parameter and $m_3$ for each one of the molecules studied in this work.

To numerically solve the eigenvalue Eq.~(\ref{282}), a slightly modified version of the Numerov method \cite{numerov, numerov2, numerov3, numerov4} will be used. The algorithm was implemented by the authors using C++ language. All the calculations were done with the CERN/ROOT package. The energies of different molecules, expressed in hartree, are shown in Table~\ref{tabelaenergia}, and the values resulting from the potential of the type $\ln(\rho)$ in two dimensions and $1/\rho$ in three dimensions were recalculated from reference~\cite{felipeepjd} with the same scales present in this paper.

\renewcommand{\arraystretch}{1.0}
\begin{table}[htb]
  \caption{The $\varepsilon$ values (in hartree) for different molecules. The first two columns represent the energy values with the Chern-Simons potential corresponding to different choices of the $\lambda$ parameter. The third column shows the values of $\varepsilon$ for logarithmic type potential and the last column represents the energies for the usual $1/\rho$ potential in three dimensions.}
  \begin{center}
  \begin{tabular}{|c|c|c|c|c|}
    \hline
      Molecule&$\varepsilon (\lambda=0.2\times10^{-3})$&$\varepsilon (\lambda=0.2\times10^{-5})$&$\varepsilon(\ln(\rho)$)&$\varepsilon(1/\rho$)           \\ \hline
      $ppe$        & 52.2987     & 5.5253   & 0.08835  & -0.6067         \\ \hline
      $dde$        & 57.9003     & 5.5117   & 0.01619  & -1.2797         \\ \hline
      $pp\mu$      & 62.9664     & 6.8929   & 18.5543  & -129.3342       \\ \hline
      $dd\mu$      & 63.6303     & 6.8760   & 3.1897   & -261.1241       \\ \hline
  \end{tabular}
  \end{center}
  \label{tabelaenergia}
\end{table}
\renewcommand{\arraystretch}{1}

\begin{figure}[htbp]
\centerline{\includegraphics[width=9.0cm]{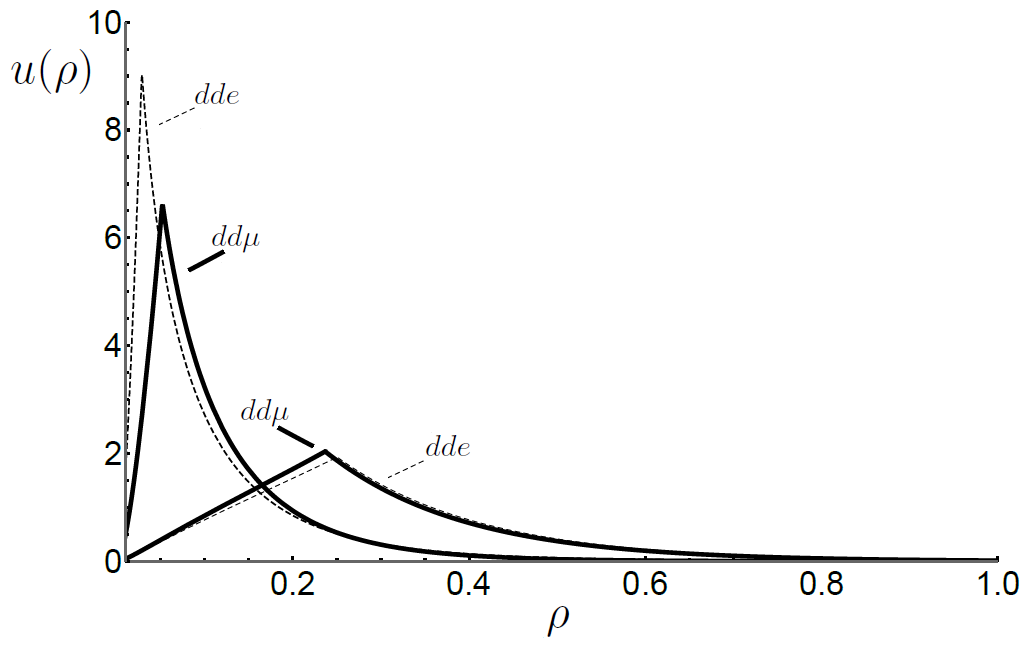}}
\caption{$dde$ and $dd\mu$ ground state wave functions for two different $\lambda$ values. The continuous curves corresponds to the choice $\lambda = 0.2 \times 10^{-5}$ and the dashed curves to $\lambda = 0.2 \times 10^{-3}$ in arbitrary units.}
\label{onda}
\end{figure}

The ground state wave functions corresponding to the four different molecules considered here are shown in figure~\ref{onda}.
\newpage

We also calculate the mean distance between the two nuclei ($\langle \rho \rangle$) inside of each molecule as shown in table~\ref{tabelmean}, using the definition
$$\langle \rho \rangle = \int_{0}^{\infty} |\psi(\rho)|^2 \rho^2 \mbox{d}\rho$$
\renewcommand{\arraystretch}{1.0}
\begin{table}[hbt]
  \caption{The $\langle \rho \rangle$ values for different molecules are given in units of Bohr's radius. The first two columns represent the $\langle \rho \rangle$ values with the Chern-Simons potential corresponding to different choice of the $\lambda$ parameter. The third column with logarithmic type potential and the last column represents the usual $1/\rho$ potential in three dimensions}
  \begin{center}
  \begin{tabular}{|c|c|c|c|c|}
    \hline
      Molecule&$(\lambda=0.2\times10^{-3})$&$(\lambda=0.2\times10^{-5})$&$ln(\rho)$&$1/\rho$           \\ \hline
      $ppe$        &  0.5308     & 0.0636  &2.0087  &1.9441             \\ \hline
      $dde$        & 0.3795      & 0.0436  &1.4332 &1.4169           \\ \hline
      $pp\mu$      & 0.4917     &  0.0537   &0.0105  &0.0101   \\ \hline
      $dd\mu$      & 0.3432      & 0.0355  &0.0076  &0.0073   \\ \hline
  \end{tabular}
  \end{center}
  \label{tabelmean}
\end{table}
\renewcommand{\arraystretch}{1}


\section{Discussions and final remarks}

In the course of this work, we were able to numerically solve four muonic and electronic molecules admitted to be governed by the Chern-Simons potential.

First of all, analyzing the results displayed in table~\ref{tabelaenergia}, we observe that the energy value for the hydrogen molecular ion ($ppe$) we have found in three dimensions (3D) is in a very good agreement with other values found in the literature~\cite{Dario, Doma}.

Only positive energies were found for the Chern-Simons potential. This is an expected fact in planar physics, in agreement with what happens in the case of $\ln(\rho)$ type interatomic interactions discussed in Ref.~\cite{felipeepjd}, where the lepton (electron or muon) is trapped in a potential well created by the nuclei.

Now, we would like to stress that the ground state energy, in the case of molecules governed by the Chern-Simons potential, depends on the choice of $\lambda$ factor, changing by a factor 10 going from the limit $\lambda \simeq 0.2 \times 10^{-3}$ to the other $\lambda \simeq 0.2 \times 10^{-5}$. This is true for all four molecules. However, for the same $\lambda$ factor, the energies hardly change between these different molecules. This indicates that the topological mass of the photon plays somehow an important role in defining the structure of these molecules. Such a behavior is qualitatively different from that shown by the molecules interacting with a potential of the $\ln(\rho)$ type in 2D, or in the usual 3D case, with $1/\rho$ potential. In both situations, indeed, we have the bound state energy of muonic molecules about 200 times smaller than that of electronic molecules, in agreement with Ref.~\cite{felipeepjd}. In addition, it should be remembered that, in the paper treating atoms governed by Chern-Simons interaction~\cite{felipe5}, it was shown that, for the Coulomb potential $1/\rho$, the bound state atomic energy and the mean atomic distance both decrease by a factor 200, when we substitute the electron by the muon in the atomic ion. Actually, in our case, the energies for $ppe$ and $dde$ molecules are quite similar, and the same occurs for $pp\mu$ and $dd\mu$. However, at the same time, the ionization energy is always bigger than the equivalent value in the case of $\ln(\rho)$ potential, which is of the order of $5\times 10$~hartree. More specifically, this energy is found to be $\simeq 100$ times (when $\lambda = 2 \times 10^{-5}$) or 200 times bigger (when $\lambda = 2 \times 10^{-3}$).

On the other side, another feature should be highlighted. We can see from table~\ref{tabelmean} that the distance between the two nuclei inside the molecules, \textit{grosso modo}, varies from a factor $\simeq 10$ going from one limit to another of the $\lambda$ factor. This mean value is indeed much smaller than the equivalent value found for the $\ln(\rho)$ potential (smaller by a factor 30 in the case of $ppe$).

So, the system governed by the Chern-Simons potential does not present an important characteristic usually attributed to muonic molecules, namely the fact that, being 200 times heavier than the electron, the muon reduces significantly the distance between nucleons as we can see in the last two columns of table~\ref{tabelmean}. In addition the ionization energy is much greater than naively expected. This fact is certainly not a good news for certain future applications, such as the study of muon catalyzed fusion \cite{felipeepl, felipeepjd} in the framework of a Chern-Simons theory. However, it is important to emphasize that within the range of the topological mass of the photon studied in this paper, the electronic molecules have an average internuclear distance significantly smaller for the Chern-Simons potential than their analogs in both 2D and 3D.

The wave functions, obtained in Figure~\ref{onda}, are similar to those found in Ref.~\cite{felipeepjd, felipe2}. They corroborate the conclusion that the topological photon mass plays an important role in system dynamics.

Finally, a comparison of the results obtained in this paper with experimental data may shed light on the real interaction that occurs in almost two-dimensional systems. Once one can perform experiments on planar materials with muonic and electronic atoms, their experimental differences can be used to rule out, or not, the presence of a Chern-Simons interaction.

\section*{Acknowledgment}

One of us (FS) was financed in part by the Coordena\c{c}\~{a}o de Aperfei\c{c}oamento de Pessoal de N\'{\i}vel Superior -- Brazil (CAPES), Finance Code 001.

\end{document}